\documentclass[conference]{IEEEtran}
\IEEEoverridecommandlockouts
\usepackage{cite}
\usepackage{amsmath,amssymb,amsfonts}
\usepackage{algorithmic}
\usepackage{graphicx}
\usepackage{textcomp}
\usepackage{xcolor}
\usepackage{verbatim}
\usepackage{booktabs}
\usepackage{url}
\usepackage{textcomp}
\usepackage{hyperref}

\def\BibTeX{{\rm B\kern-.05em{\sc i\kern-.025em b}\kern-.08em
    T\kern-.1667em\lower.7ex\hbox{E}\kern-.125emX}}
\begin{document}

\title{Predicting Gene Disease Associations in type 2 Diabetes Using Machine Learning on Single-Cell RNA-Seq Data\\
}
\author{
\IEEEauthorblockN{1\textsuperscript{st} Daniel F.~O.~Onah}
\IEEEauthorblockA{
\textit{Department of Information Studies} \\
\textit{University College London} \\
London, United Kingdom \\
d.onah@ucl.ac.uk
}
\and
\IEEEauthorblockN{2\textsuperscript{nd} Maria-de la Luz Lomboy Toledo}
\IEEEauthorblockA{
\textit{Department of Information Studies} \\
\textit{University College London} \\
London, United Kingdom \\
maria.toledo.24@ucl.ac.uk
}
}

\maketitle

\begin{abstract}
Diabetes is a chronic metabolic disorder characterized by elevated blood glucose levels due to impaired insulin production or function. Two main forms are recognized: type~1 diabetes (T1D), which involves autoimmune destruction of insulin-producing $\beta$-cells, and type~2 diabetes (T2D), which arises from insulin resistance and progressive $\beta$-cell dysfunction. Understanding the molecular mechanisms underlying these diseases is essential for the development of improved therapeutic strategies, particularly those targeting $\beta$-cell dysfunction.

To investigate these mechanisms in a controlled and biologically interpretable setting, mouse models have played a central role in diabetes research. Owing to their genetic and physiological similarity to humans, together with the ability to precisely manipulate their genome, mice enable detailed investigation of disease progression and gene function. In particular, mouse models have provided critical insights into $\beta$-cell development, cellular heterogeneity, and functional failure under diabetic conditions.

Building on these experimental advances, this study applies machine learning methods to single-cell transcriptomic data from mouse pancreatic islets. Specifically, we evaluate two supervised approaches identified in the literature; Extra Trees Classifier (ETC) and Partial Least Squares Discriminant Analysis (PLS-DA), to assess their ability to identify T2D-associated gene expression signatures at single-cell resolution. Model performance is evaluated using standard classification metrics, with an emphasis on interpretability and biological relevance.
\end{abstract}


\section{Introduction}

Advances in mouse genomics have greatly facilitated the study of type~2 diabetes (T2D) at the gene-expression level. In particular, single-cell RNA sequencing (scRNA-seq) technologies enable high-resolution profiling of individual pancreatic islet cells, capturing cell-to-cell variability that is masked in bulk analyses. Leveraging these advances, Hrovatin \emph{et al.}~\cite{b1} compiled a comprehensive cross-condition Mouse Islet Atlas (MIA), integrating over 300,000 single cells from 56 samples across nine independent datasets. These datasets span a wide range of biological contexts, including developmental stages from embryonic to aged mice, both sexes, and multiple disease states. Importantly, the atlas incorporates several well-established diabetes models, such as the autoimmune Non-Obese Diabetic (NOD) mouse for type\textasciitilde 1 diabetes, the db/db model for type\textasciitilde 2 diabetes, and the $\beta$-cell ablation model (STZ).

By combining data from healthy, immature, aged, and diseased $\beta$--cells, the Mouse Islet Atlas (MIA) provides a unified view of how these cells change under different conditions. This approach has identified new $\beta$--cell states that appear during disease and revealed key molecular pathways related to stress, loss of cell identity, and compensation. In particular, $\beta$--cells from the db/db model, a mouse model for type\textasciitilde 2 diabetes caused by a mutation in the leptin receptor gene, which leads to obesity, insulin resistance, and progressive $\beta$--cell failure, showed gene expression patterns very similar to those in human T2D $\beta$--cells. This supports the relevance of the db/db model for studying the molecular basis of human diabetes\cite{b1}.

While the Mouse Islet Atlas provides a comprehensive and unified view of $\beta$--cell states across diverse biological conditions, the analysis of such scRNA-seq data remains inherently challenging. Single-cell transcriptomic datasets are highly dimensional and subject to technical noise, biological variability, and batch effects, particularly when integrating data from multiple studies \cite{b2}. Although standard analytical approaches, including clustering and differential gene expression analysis, are effective for defining major cell populations and identifying marker genes, they often offer a limited representation of the complex transcriptional relationships underlying disease-associated cellular states \cite{b3}. These challenges highlight the need for integrative computational frameworks capable of systematically leveraging large-scale scRNA-seq resources to better characterize $\beta$--cell heterogeneity and dysfunction in type~2 diabetes.

In recent years, advances in computational biology have begun to bridge this gap. Machine learning (ML) techniques provide a powerful framework for uncovering latent patterns and generating predictions from transcriptomic data. Supervised ML models can be trained to classify healthy versus diabetic cell states based on gene expression signatures, whereas unsupervised approaches can identify previously unknown cell subtypes or transitional cellular states. The availability of large, curated resources such as the Mouse Islet Atlas enables the application of these data-intensive methods with increased robustness and biological interpretability. Thus, integrating machine learning with scRNA-seq data holds significant potential for advancing our understanding of $\beta$-cell dysfunction and accelerating the discovery of therapeutic targets in type~2 diabetes.

This gap has motivated the application of machine learning (ML) techniques in the domain of gene--disease association prediction. ML algorithms can analyze complex multivariate patterns in gene expression data and learn classifiers or predictors that map expression profiles to a disease outcome (e.g., diabetic or normal). By training on examples of known diabetic vs.\ non-diabetic cell profiles, an ML model can potentially generalize to identify new cases or key disease-associated genes.

Recent work by Li \emph{et al.}~\cite{b4} demonstrates the power of this approach: they analyzed single-cell RNA-seq data from human islet cells (949 T2D and 651 control cells) using several machine-learning (ML) algorithms and feature selection methods, including Support Vector Machines (SVM), Random Forest (RF), Extreme Gradient Boosting (XGBoost), and LASSO-based feature selection. This analysis discovered novel T2D-associated genes (e.g., MTND4P24, MTND2P28) and even inferred logical rules involving multiple genes that could distinguish diabetic from healthy cells. These insights were obtained by going beyond single-gene statistics to a multivariate ML framework. Such findings underscore that ML can reveal complex gene signatures of disease that conventional analyses might miss.

\subsection*{Research Objectives}

This study will use machine learning techniques to improve the identification of diabetic and non-diabetic $\beta$-cells based on their gene expression signatures. By learning from complex, high-dimensional transcriptomic data, these models aim to detect non-obvious patterns that may be indicative of disease presence, supporting both predictive accuracy and biological interpretability.

This provides a scalable and data-driven approach to disease classification. This dual approach seeks not only to assess predictive performance, but also to contribute to a deeper understanding of the molecular underpinnings of T2D and to identify potential gene biomarkers for future research.

\subsection*{Methodology}

First, literature review is conducted to establish the theoretical and scientific foundation for this study. This review covers the use of mouse models in diabetes and genomics research, transcriptomic analysis and advances in RNA sequencing technologies, particularly single-cell RNA sequencing (scRNA-seq), and the application of machine learning (ML) methods to classify or predict type~2 diabetes (T2D) based on gene expression data.

Then, a data analysis is performed, focused on extracting and preparing scRNA-seq data from the Mouse Islet Atlas, specifically the db/db and mSTZ samples, for type~2 diabetes. The original h5ad-format files are loaded and processed using dedicated Python libraries such as \texttt{scanpy} and \texttt{anndata}. Cells are filtered to retain only pancreatic $\beta$-cells, and standard preprocessing steps are applied, including normalization, log-transformation, and feature selection (e.g., highly variable genes). The resulting dataset is reduced in size to ensure efficient computation on the local system.

Finally, machine learning models are implemented to classify cells as diabetic or non-diabetic based on their gene expression profiles. The classification pipeline is developed in Python, with dimensionality reduction and resampling to balance classes. Performance is evaluated using cross-validation and standard metrics such as accuracy, precision, recall, F1-score, and AUC-ROC. This integrated approach enables the identification of transcriptional patterns associated with T2D and the assessment of ML as a diagnostic tool in single-cell genomics.

\section{Literature Review}

\subsection{Diabetes as a Disease}

Diabetes mellitus is a group of metabolic disorders characterized by chronic hyperglycaemia resulting from impaired insulin secretion, insulin action, or both. The two main forms are type~1 diabetes (T1D) and type~2 diabetes (T2D). T1D is primarily associated with autoimmune-mediated destruction of pancreatic $\beta$--cells, leading to an absolute deficiency of insulin, whereas T2D is characterized by insulin resistance accompanied by progressive $\beta$--cell dysfunction and insufficient insulin secretion \cite{b4}. As insulin demand increases, $\beta$--cells initially attempt to compensate, but this adaptive capacity declines over time, resulting in sustained hyperglycaemia.

Type~2 diabetes is the predominant form of the disease, accounting for approximately 90\% of all diabetes cases worldwide, and is strongly associated with obesity and environmental factors such as physical inactivity and sedentary lifestyle \cite{b4,b5}. These factors impose chronic metabolic stress on pancreatic islet cells, contributing to impaired insulin sensitivity and gradual loss of $\beta$--cell function. Consequently, T2D is increasingly recognized as a complex and multifactorial disease involving both systemic metabolic disturbances and intrinsic alterations in pancreatic islet biology.

Diabetes has emerged as a major global public health challenge. In 2015, an estimated 415~million people were affected worldwide, and this number is projected to rise to 642~million by 2040, driven largely by the growing prevalence of T2D \cite{b5}. These trends underscore the urgent need to better understand the cellular and molecular mechanisms underlying $\beta$--cell dysfunction in order to support the development of more effective preventive and therapeutic strategies.

\subsection{Genetic Factors and Implicated Genes in Diabetes}

Type~2 diabetes is a genetically complex disease in which susceptibility arises from the combined effects of numerous genetic variants, each contributing modestly to disease risk. Family and population studies indicate a strong heritable component; however, the genetic architecture of T2D is highly polygenic and shaped by interactions among multiple genes and environmental factors \cite{b9}. As a result, identifying causal genes and mechanisms has proven challenging, as many associated variants exert small effects and act within interconnected biological networks rather than isolated pathways.

Large-scale genetic studies have revealed that a substantial proportion of T2D-associated variants are linked to pancreatic islet function, particularly pathways regulating insulin secretion from $\beta$--cells \cite{b8}. Many risk variants are located in non-coding regions of the genome, suggesting that altered gene regulation, rather than changes in protein sequence, plays a key role in disease development. These findings support the view that impaired $\beta$--cell function is a central component of T2D pathogenesis and highlight the need for integrative approaches that combine genetic variation with molecular and cellular phenotypes to better understand disease mechanisms \cite{b8,b9}.

In summary, T2D risk is mediated by multiple genetic variants that primarily act by subtly modulating $\beta$-cell function and related metabolic pathways. As genomic and analytical tools continue to advance, additional implicated genes are likely to be identified, providing promising targets for future therapeutic development and precision medicine approaches in the future.

\subsection{The Mouse Islet Atlas for Diabetes Analysis}

Hrovatin \emph{et al.}~\cite{b1} developed the Mouse Islet Atlas (MIA), a unified single-cell transcriptomic resource that integrates more than 300{,}000 cells across healthy and diabetic conditions. For this study, MIA is important because it provides a high-quality, well-annotated reference of $\beta$-cell states that captures how gene expression changes during diabetes. By harmonizing data from multiple diabetic models, MIA reveals which transcriptional signatures are robust and consistently associated with $\beta$-cell dysfunction.

Moreover, the atlas shows that $\beta$-cells under metabolic or toxin-induced stress (db/db and STZ models) share strong transcriptional similarities with human T2D $\beta$-cells. This offers biologically relevant ground truth for distinguishing diabetic from healthy cells based solely on gene expression, which is the aim of the present analysis. MIA also highlights key pathways altered in diabetes, such as stress-response programs and loss of $\beta$-cell maturity markers, providing biological context for the gene-level signals later identified by the machine-learning models.

In summary, the Mouse Islet Atlas offers the foundational transcriptomic landscape that enables this study to explore how single-cell gene expression profiles discriminate healthy from diabetic $\beta$-cells and to evaluate whether machine-learning models can capture these disease-associated patterns with high accuracy.

\subsection{Machine Learning for Genomic and Transcriptomic Analysis}

The rapid growth of high-throughput genomic technologies in diabetes research has made machine learning (ML) a vital tool for uncovering patterns in large and complex datasets. ML algorithms are especially useful for identifying relevant features, such as genes, SNPs, and methylation sites, associated with disease and for building predictive models of molecular phenotypes that go beyond traditional statistical methods.

One major application is the integration of multi-omics, epigenomic, and transcriptomic information to better understand gene regulation in diabetes. Rather than analysing each data type separately, ML methods can link them to identify shared biomarkers. For example, R{\"o}nn \emph{et al.} applied a supervised ML method called DIABLO (Data Integration Analysis for Biomarker discovery using Latent Components), based on partial least squares (PLS), to human pancreatic islet data from 110 donors, approximately 30\% of whom had T2D. By integrating RNA-seq, DNA methylation, SNP genotypes, and clinical data, the model achieved high classification performance, with an accuracy of approximately 91\% and an AUC of 0.96, in distinguishing T2D from control islets~\cite{b8}.

Importantly, the same study also evaluated a simpler ML approach using only RNA-seq data in combination with a PLS-DA (Partial Least Squares Discriminant Analysis) model. Using a subset of 38 selected genes, this model achieved a classification accuracy of 84--90\% and an AUC of 0.92--0.96, performance levels comparable to those of the full multi-omics model. These results demonstrate that even when restricted to transcriptomic data alone, supervised ML methods can identify robust gene expression signatures associated with T2D.

A deeper examination of PLS-DA illustrates why this method has become increasingly popular in genomic and transcriptomic studies. As Mehmood \emph{et al.} emphasize, PLS-DA is particularly well suited for high-dimensional biological data, where the number of variables, such as genes, greatly exceeds the number of samples. By projecting predictors into a latent space that maximizes covariance with the response variable, PLS-DA effectively reduces dimensionality while preserving discriminative information ~\cite{b10}. 

The Mouse Islet Atlas provides a framework to study these phenomena by enabling direct comparison of gene expression profiles from immature, mature, and failing $\beta$-cells. For instance, pathways related to oxidative stress responses or fetal developmental programs may be reactivated in ageing or diabetic $\beta$-cells, suggesting a reversion to progenitor-like features, a form of de-differentiation observed in diabetic islets. Hrovatin \emph{et al.} also introduced the concept of $\beta$-cell ``compensation'' versus ``failure.'' In a T2D context, $\beta$-cells initially compensate for insulin resistance by proliferating and increasing insulin secretion, but this compensatory capacity eventually fails ~\cite{b1}. In this setting, PLS-DA not only reduces dimensionality but also enhances the interpretability of discriminant features, making it particularly suitable for transcriptomic datasets in which correlated gene expression patterns may obscure biologically relevant signals. In genomic research, PLS-DA has been widely applied to classify disease subtypes, identify biomarker panels, and highlight molecular pathways driving group separation. Compared with traditional methods, its ability to handle multicollinearity and noise while retaining predictive accuracy positions it as a powerful supervised learning tool for uncovering disease-associated expression signatures in diabetes and related conditions~\cite{b10}.

In summary, machine learning provides powerful tools for analysing genomic and transcriptomic data in diabetes research. It enables the integration of heterogeneous data types, the discovery of complex biological patterns, and the identification of candidate biomarkers that can subsequently be validated experimentally. As demonstrated by studies such as that of R{\"o}nn \emph{et al.}, ML is instrumental in revealing multilayer disease signatures, even when applied to a single data modality such as RNA-seq.

\subsection{Machine Learning for Diabetes Prediction}

Machine learning (ML) is widely applied in diabetes research not only to analyse genetic and transcriptomic data but also to build predictive models that support early diagnosis and risk stratification. Diabetes is a multifactorial disease influenced by genetic, clinical, demographic, and lifestyle factors. This complexity makes it well suited to ML approaches capable of handling non-linear relationships and high-dimensional data.

Machine learning models can integrate dozens or even hundreds of features, thereby improving prediction accuracy and enabling more nuanced insights. Commonly used algorithms include logistic regression, decision trees, ensemble methods such as random forests and gradient boosting, support vector machines, and neural networks. These models are typically trained on datasets containing variables such as age, body mass index (BMI), blood pressure, glucose levels, and family history, with the aim of predicting either current diabetic status or future disease onset.

An illustrative example is the study by De \emph{et al.}, who evaluated how gut microbiome profiles, combined with physiological and biochemical measurements, could improve classification of T2D. Their ML models confirmed established predictors such as HbA1c and fasting glucose, while also identifying additional signals, including microbial abundances and lipid ratios. These findings point toward a future in which ML-based risk models integrate clinical, molecular, and microbiome data to achieve more accurate and personalized prediction~\cite{b5}.

Another notable example is the work by Hama~Saeed \emph{et al.}, which focused on classifying type~2 diabetes using only publicly available clinical datasets. The authors employed several ML classifiers, including Random Forest, Decision Tree, and Extra Trees Classifier (ETC), in combination with oversampling techniques to address class imbalance. Among the models evaluated, the Extra Trees Classifier achieved the best performance, with an accuracy of 95\% and an F1-score of 0.95. The ETC is an ensemble learning method that constructs multiple uncorrelated decision trees using random subsets of features and samples, providing both robustness and interpretability. The model also identified fasting glucose, BMI, age, and insulin levels as the most influential predictors for T2D, illustrating how ML can rank feature importance and highlight key risk factors~\cite{b11}.

The Extra Trees Classifier (ETC) is particularly well suited for high-dimensional biomedical datasets due to its strong randomization strategy during tree construction. Unlike Random Forests, ETC builds each tree using the full training dataset and selects split thresholds at random, which reduces variance, limits overfitting, and improves computational efficiency \cite{b21}. In addition to its classification performance, ETC provides an effective embedded feature selection mechanism through impurity-based importance measures, allowing relevant variables to be ranked according to their contribution to the model.

In conclusion, machine learning provides powerful tools for improving diabetes prediction and diagnosis. ML models can integrate heterogeneous inputs, identify key predictors, and outperform traditional statistical approaches in many scenarios. Among candidate methods, the Extra Trees Classifier stands out as a robust, variance-reduced, and interpretable baseline for high-dimensional settings typical of genomics and transcriptomics. In multimodal contexts, combining clinical measurements with molecular or microbiome features yields additional predictive signals, as illustrated by De \emph{et al.} and Hama~Saeed \emph{et al.}. Furthermore, ensemble methods such as ETC not only enhance predictive accuracy but also provide interpretable feature rankings, enabling clinicians to better understand the factors driving model decisions. As data availability and model transparency continue to improve, ML-based prediction tools are likely to become increasingly important in personalized diabetes care.

\section{Methodology}

\begin{figure}
    \centering
    \includegraphics[width=1\linewidth]{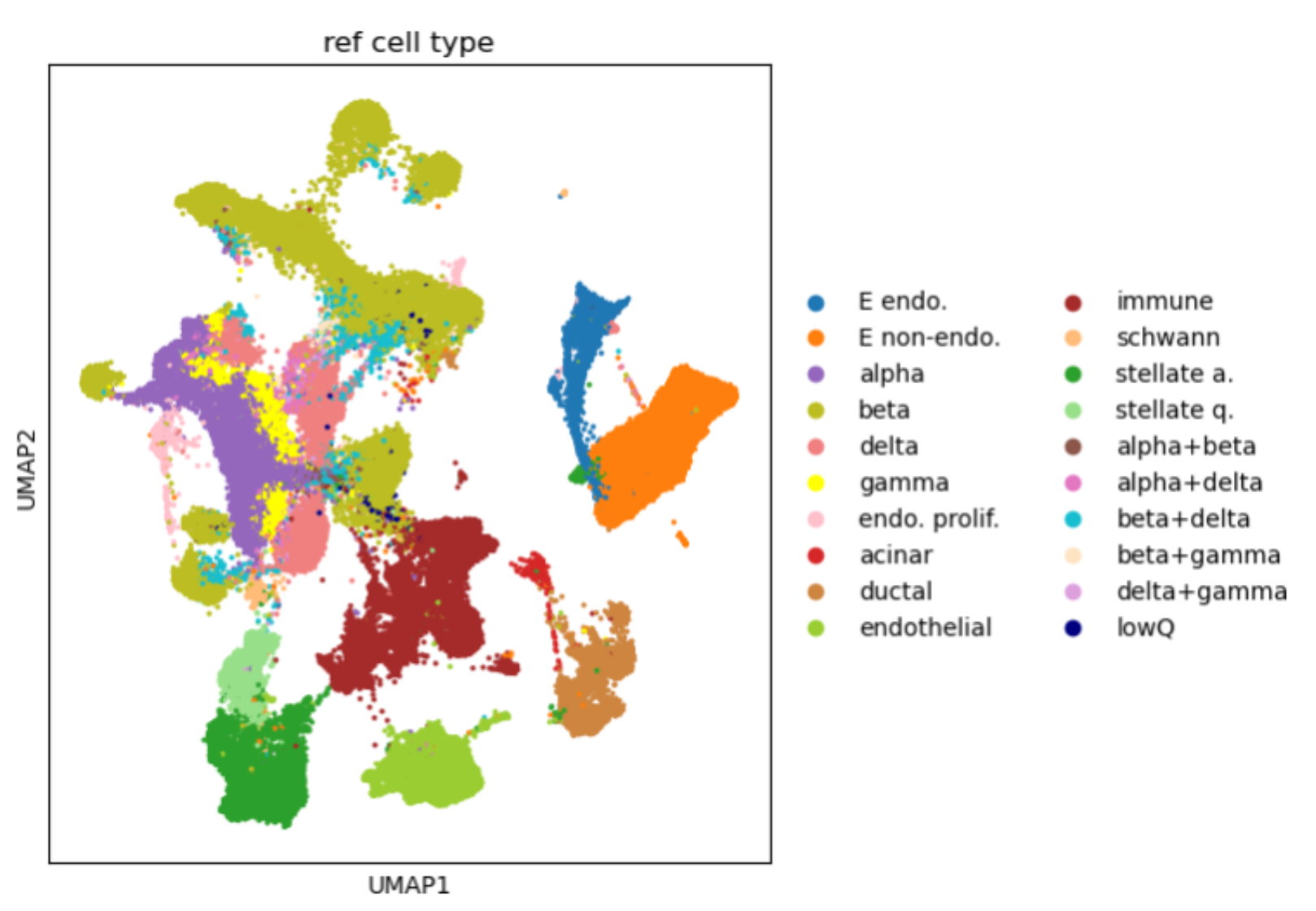}
    \caption{UMAP projection of annotated cell types in the Mouse Islet Atlas. Each colour denotes a specific cell type, including endocrine populations (e.g., $\alpha$, $\beta$, $\delta$), exocrine cells (acinar and ductal), and non-pancreatic cell types such as immune, endothelial, and stellate cells.}
    \label{fig:umap_celltypes}
\end{figure}

The methodology begins with the preprocessing and analysis of single-cell RNA sequencing (scRNA-seq) data from the Mouse Islet Atlas (MIA), as shown in Fig.~\ref{fig:umap_celltypes}. Following an initial inspection of the available datasets, the analysis is restricted to adult samples in order to focus on mature pancreatic tissue. Within this subset, only cells annotated as normal (healthy) or type~2 diabetic (T2D) are retained.

Consistent with the literature reviewed in Section~II, downstream analyses are further limited to pancreatic $\beta$-cells. This filtering ensures that the machine learning models operate on a biologically relevant and homogeneous cell population, avoiding confounding effects driven by differences in cell-type composition. Data filtering and handling are performed using the Python libraries \texttt{scanpy} and \texttt{anndata}, operating on the original \texttt{.h5ad} files.

Standard preprocessing steps are then applied to the resulting $\beta$-cell expression matrix. These include normalization, log-transformation, and the selection of highly variable genes, preparing the data for downstream machine learning analysis while ensuring computational efficiency.\footnotemark

\footnotetext{Code available at: \href{https://github.com/mdelaluzl/t2d-gene-disease-ml-prediction}{github.com/mdelaluzl/t2d-gene-disease-ml-prediction}.}

\subsection*{Code Availability}
The full implementation of the preprocessing pipeline, machine learning models (PLS-DA and Extra Trees Classifier), and evaluation scripts used in this study is publicly available on GitHub at:
\href{https://github.com/mdelaluzl/t2d-gene-disease-ml-prediction}{https://github.com/mdelaluzl/t2d-gene-disease-ml-prediction}.

\subsection*{Machine-Learning Models Evaluated}

We tested two machine learning models that are well suited for high-dimensional biological data, Partial Least Squares Discriminant Analysis (PLS-DA) and Extra Trees Classifier (ETC). We begin with a description of the overall method, evaluating performance models, followed by the methodology used.

\subsubsection*{A. Partial Least Squares Discriminant Analysis (PLS-DA)}

PLS-DA is a supervised extension of partial least squares that constructs a small number of latent components from the high-dimensional gene matrix $\mathbf{X}$ so that these components capture as much covariance as possible with an encoded class vector or matrix $\mathbf{Y}$ (diabetic vs.\ control). In practical terms, the method learns new axes in the gene expression space along which the two groups differ most. Each cell is projected onto these axes to obtain scores, and classification is performed in this reduced space.

Two properties highlighted by Mehmood \emph{et al.}~\cite{b10} make PLS-DA attractive for scRNA-seq settings with $p \gg n$:
\begin{enumerate}[(i)]
\item it handles strong multicollinearity among genes by modelling shared variation through a few supervised components; and
\item it yields transparent gene-level interpretation via component loadings and Variable Importance in Projection (VIP) scores that rank features by their contribution to the discriminative components.
\end{enumerate}

We adopt the VIP definition of Mahieu \emph{et al.}~\cite{b13} for PLS1:

\begin{equation}
\mathrm{VIP}_j =
\sqrt{
p \,
\frac{
\sum_{h=1}^{m} R^2(y, t_h)\, w_{hj}^2
}{
\sum_{h=1}^{m} R^2(y, t_h)
}
},
\end{equation}

where $p$ is the number of genes, $m$ the number of retained components, $R^2(y, t_h)$ the proportion of response variance explained by component $h$, and $w_{hj}$ the loading weight of gene $j$ on component $h$. For multivariate $\mathbf{Y}$ (PLS2), we follow the matrix/trace generalization described in Section~2.1 of Mahieu \emph{et al.}~\cite{b13}.

\subsubsection*{B. Extra Trees Classifier (ETC)}

The Extra Trees Classifier (ETC) is an ensemble method that constructs a forest of highly randomized decision trees for supervised classification tasks. Unlike Random Forests, which rely on bootstrap sampling and search for near-optimal split thresholds within random feature subsets, ETC typically grows each tree on the full training set and introduces stronger randomization by selecting both the candidate features and their cut-points at random before choosing the best split~\cite{b14,b15}. This additional level of randomness reduces variance and helps control overfitting, providing competitive accuracy with improved computational efficiency~\cite{b14,b15}.

Importantly, such randomized splits are particularly advantageous in high-dimensional and noisy datasets, as they prevent the model from overfitting spurious patterns. In biomedical contexts, this property makes ETC especially well suited for omics and clinical data. For example, when applied to type~2 diabetes classification tasks, ETC achieved superior ROC/AUC performance compared with other tree-based models~\cite{b11}. Therefore, the method's strong randomization not only enhances robustness against noisy biological signals but also improves interpretability through feature importance analysis.

In Extra Trees, feature importance is measured by the Mean Decrease in Impurity (MDI). For a feature $j$, MDI is defined as the average reduction in node impurity whenever $j$ is used for a split, weighted by the proportion of samples reaching the node, and normalized such that $\sum_j \mathrm{MDI}_j = 1$:

\begin{equation}
\mathrm{MDI}_j =
\frac{
\sum_{t=1}^{T} \sum_{n \in \mathcal{N}_t} \mathbb{I}\{f(n)=j\}\, p(n)\, \Delta i(n)
}{
\sum_{k} \sum_{t=1}^{T} \sum_{n \in \mathcal{N}_t} \mathbb{I}\{f(n)=k\}\, p(n)\, \Delta i(n)
},
\end{equation}

where $\Delta i(n) = i(n) - p_L\, i(n_L) - p_R\, i(n_R)$ and $p(n)=N_n/N$ denotes the proportion of samples at node $n$. Here $i(\cdot)$ represents an impurity measure, such as the Gini index or entropy.

This criterion is fast, embedded, and widely used for variable screening. Its theoretical properties in randomized trees support its application in high-dimensional omics settings~\cite{b16,b17}.

\subsection*{Evaluating performance}

Evaluating the performance of classification models is essential, particularly in genomics and biomedical applications where data are complex and often imbalanced. Commonly used metrics include accuracy, precision (positive predictive value, PPV), recall (sensitivity), F1-score, and AUC--ROC. Each metric captures a different aspect of model quality, and the appropriate choice depends on the data distribution and the study objective~\cite{b18}. Below, we summarize what each metric measures, typical use cases in genomic contexts, and key caveats such as class imbalance and other sources of bias.

\subsubsection*{A. Accuracy}

Accuracy is defined as the proportion of correctly classified instances. It is simple to interpret and widely reported; however, it can be misleading in the presence of class imbalance, for example when control samples greatly outnumber disease cases. In such scenarios, high accuracy may be achieved by predicting only the majority class. Therefore, accuracy should not be used in isolation when the minority class is of primary interest~\cite{b18}.

\subsubsection*{B. Precision and Recall}

Precision, $\mathrm{Prec} = \mathrm{TP}/(\mathrm{TP}+\mathrm{FP})$, quantifies the reliability of positive predictions, whereas recall (sensitivity), $\mathrm{Rec} = \mathrm{TP}/(\mathrm{TP}+\mathrm{FN})$, measures the ability to capture true positives.

These metrics involve different trade-offs: increasing recall often lowers precision, resulting in more positives being identified but with a higher rate of false alarms, and vice versa.

Their use originates in information retrieval and has become standard in machine learning due to their dependence on the decision threshold and their direct interpretability for rare events. Because the choice of threshold shifts this trade-off, precision and recall should be interpreted jointly at clinically meaningful cut-offs~\cite{b18,b19}.

\subsubsection*{C. F$_1$-score}

The F$_1$-score is defined as the harmonic mean of precision and recall. It provides a single summary measure that is high only when both precision and recall are high. The F$_1$-score is particularly useful when classes are imbalanced and the study seeks a balance between avoiding false alarms and missing true cases. Because the F$_1$-score ignores true negatives, it is especially appropriate when the negative class is abundant and the primary interest lies in detecting positives. Reporting the F$_1$-score alongside its individual components enhances interpretability under class imbalance~\cite{b18}.

\subsubsection*{D. AUC--ROC}

The area under the receiver operating characteristic (ROC) curve summarizes model performance across all possible classification thresholds by integrating the trade-off between the true positive rate and the false positive rate~\cite{b18,b19}. AUC--ROC is convenient for model comparison; however, it does not encode the relative costs of false positives versus false negatives and may appear overly optimistic in highly imbalanced datasets. For practical applications, AUC--ROC should therefore be complemented with threshold-specific metrics, such as precision, recall, and F$_1$-score, evaluated at a chosen operating point~\cite{b18}.

\section{Findings}

\subsection*{Findings from PLS-DA: Discriminating Diabetic vs.\ Healthy Profiles from Gene Expression Data}

To explore transcriptomic differences between diabetic and non-diabetic individuals, we applied a supervised machine learning approach known as Partial Least Squares Discriminant Analysis (PLS-DA). As previously mentioned, PLS-DA operates by projecting the original high-dimensional gene expression data into a smaller number of latent components that maximize the separation between predefined groups. In this case, the groups correspond to healthy individuals and those with type~2 diabetes (T2D).

\subsubsection*{A. Model training and prediction}

The model was trained on a gene expression matrix, where each row represents a single-cell transcriptomic profile and each column corresponds to the expression level of a particular gene. Labels indicating whether each cell originated from a diabetic or healthy donor were used to guide the training process.

The trained PLS-DA model was then evaluated on a separate test set comprising more than 13{,}000 single-cell profiles.

\subsubsection*{B. Classification performance}

\begin{figure}
    \centering
    \includegraphics[width=1\linewidth]{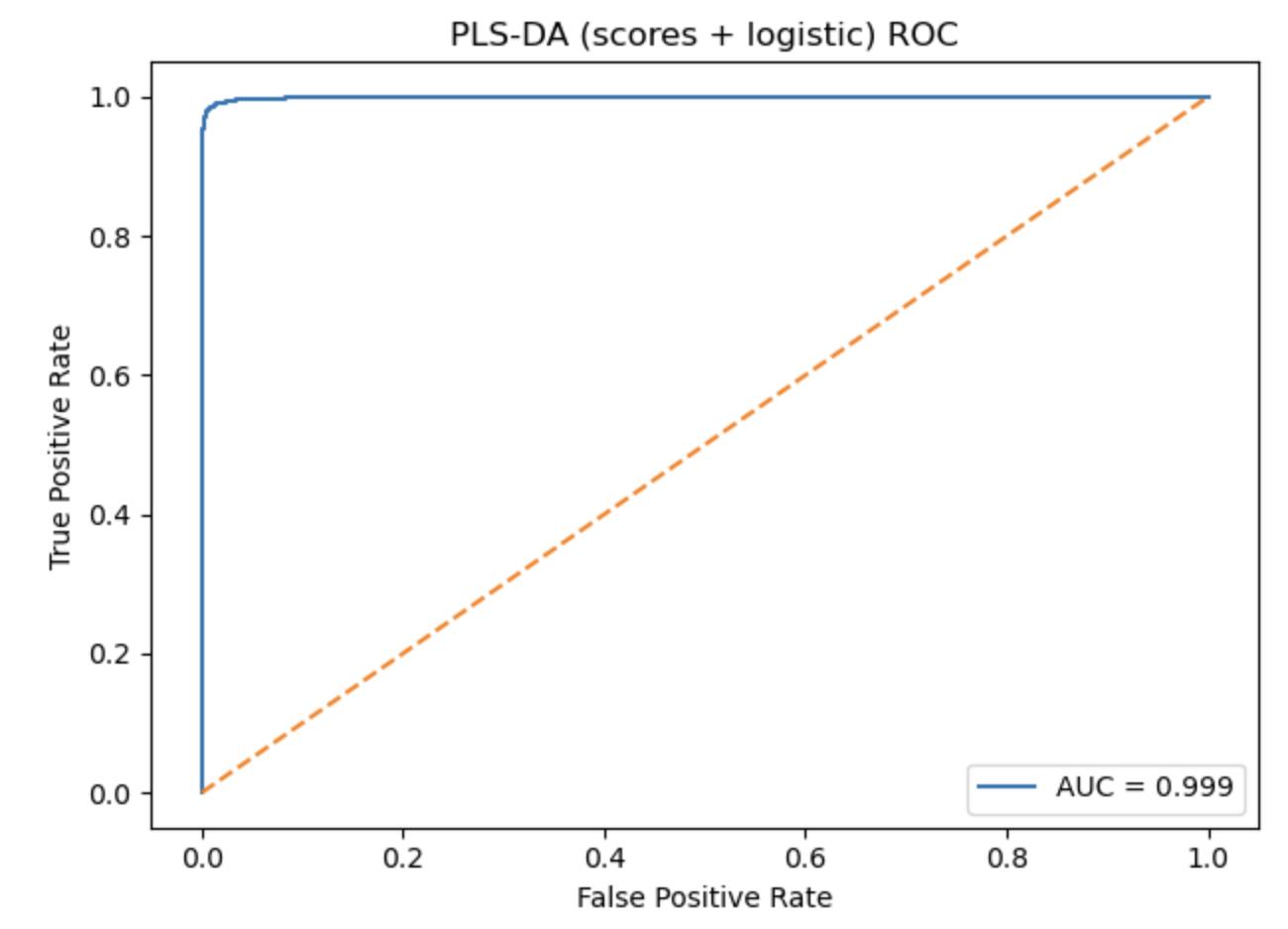}
    \caption{ROC curve for the PLS-DA model evaluated on the test set. The area under the curve (AUC) is 0.999, indicating near-perfect discrimination between diabetic and healthy cells.}
    \label{fig:plsda_roc}
\end{figure}

The model achieved exceptional classification accuracy, with an overall accuracy of 98.9\%, precision of 98.8\%, recall of 98.5\%, and an F$_1$-score of 98.6\%. These metrics reflect the model's ability to correctly distinguish diabetic from healthy cells. The high AUC--ROC value of 0.999 (Fig.~\ref{fig:plsda_roc}) indicates near-perfect discriminative power between the two conditions. Similarly high discriminative performance for PLS-DA has been reported in independent diabetes studies using alternative data modalities. For instance, Furman et al.\cite{b23} applied PLS-DA to FTIR-MIR serum spectra and achieved classification accuracies above 95\% for diabetes diagnosis, supporting the robustness of PLS-DA in capturing diabetes-associated molecular signatures beyond transcriptomic data.

\subsubsection*{C. Understanding misclassifications}

\begin{figure}
    \centering
    \includegraphics[width=1\linewidth]{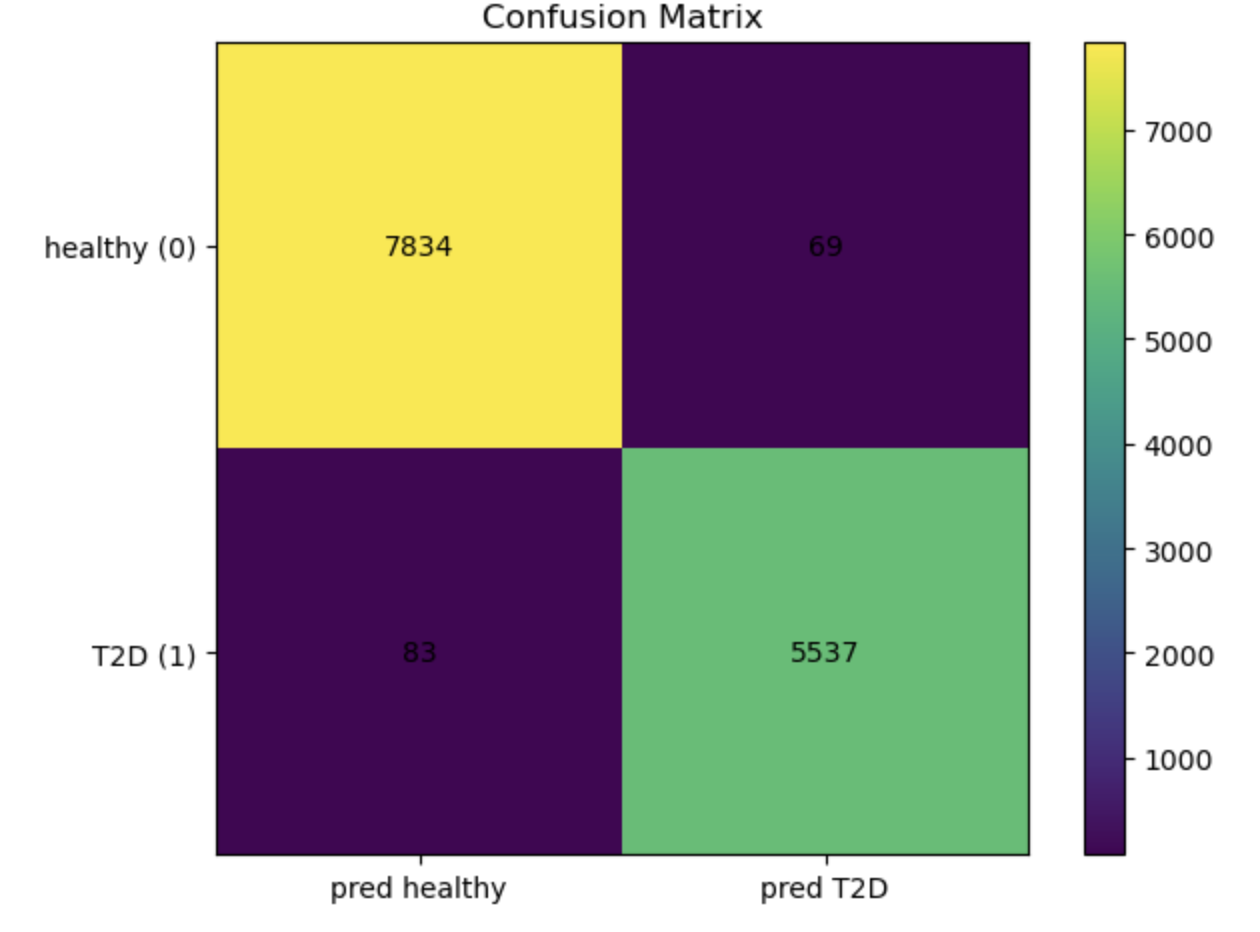}
    \caption{Confusion matrix summarizing the PLS-DA model predictions on the test set. Rows correspond to true labels, and columns correspond to predicted labels. Correct classifications lie along the diagonal.}
    \label{fig:confusion_matrix}
\end{figure}

The confusion matrix (Fig.~\ref{fig:confusion_matrix}) shows that among 13{,}523 test samples, only 152 were misclassified. Specifically, 69 healthy cells were incorrectly labelled as diabetic, while 83 diabetic cells were misclassified as healthy. This low error rate demonstrates the reliability of the model in distinguishing disease states based on gene expression profiles.

\subsubsection*{D. Latent space representation}

\begin{figure}
    \centering
    \includegraphics[width=1\linewidth]{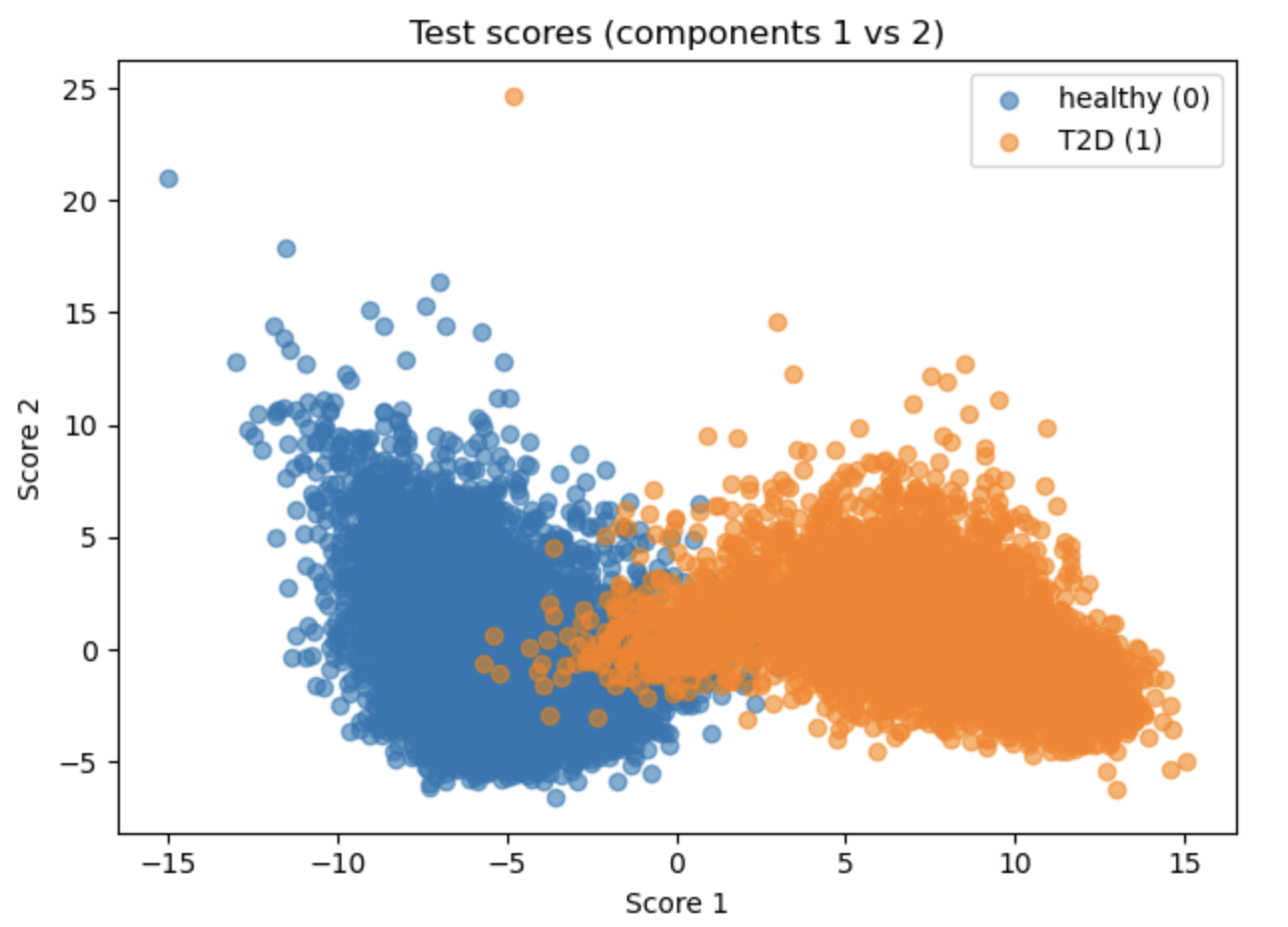}
    \caption{PLS-DA scores for test samples projected onto the first two latent components. Blue points correspond to healthy cells, while orange points correspond to T2D cells. The clear separation between groups indicates strong transcriptomic differences between conditions.}
    \label{fig:plsda_latent}
\end{figure}

The separation achieved by the model is further visualized in the scatter plot of the first two PLS-DA components (Fig.~\ref{fig:plsda_latent}). Each point corresponds to a cell, and the axes represent new synthetic dimensions learned by the model to best separate healthy and diabetic samples. The distinct clustering of orange (T2D) and blue (healthy) points confirms that gene expression patterns alone carry strong information about disease status.

\subsubsection*{E. Gene importance and biological interpretation}

To identify which genes were most influential in separating diabetic from healthy $\beta$-cells, the PLS-DA model produced a ranking based on Variable Importance in Projection (VIP) scores. In practical terms, this ranking indicates which genes carry the strongest signal for distinguishing between the two groups. The highest-ranked genes are reported in Appendix~I.

Several of these genes are associated with processes known to be relevant for $\beta$-cell function. For example, members of the secretogranin family (II, III, and V) are involved in the formation of secretory vesicles, which play a key role in insulin storage and release. Their prominence among the top-ranked features suggests that pathways related to insulin packaging and secretion are altered in diabetic cells. Other genes, such as metallothionein~2, are linked to cellular stress responses, consistent with evidence that diabetic $\beta$-cells experience increased oxidative and metabolic stress. The appearance of the vitamin~D binding protein among the most influential features further points to alterations in pathways connecting metabolism, immune regulation, and $\beta$-cell health. Additional genes, including cholecystokinin and CD81, are related to endocrine signalling and cell-to-cell communication, indicating broader disruptions in $\beta$-cell regulatory networks.

At a higher level, these results confirm that the genes most important for distinguishing diabetic from healthy cells cluster around three main biological themes:
\begin{enumerate}
\item insulin secretion;
\item stress and survival mechanisms; and
\item metabolic and immune regulation.
\end{enumerate}

This supports the interpretation that type\textasciitilde 2 diabetes in $\beta$-cells is not caused by a single gene defect, but by a combination of disrupted insulin release, heightened stress, and altered signalling pathways.

\subsubsection*{F. Conclusion}

Overall, this PLS-DA analysis demonstrates that gene expression signatures can be used with high accuracy to distinguish between diabetic and healthy states at the single-cell level. Considering the near-perfect AUC, balanced confusion matrix, and clear latent-space separation support that the discriminant signal is strong and not driven by a single class. In that sense, the model not only performed well in classification but also provides biologically meaningful gene rankings that may guide further studies in biomarker discovery and disease finding.

Finally, the VIP-derived gene set offers biologically interpretable leads that align with endocrine secretion and cellular stress pathways, making it a reasonable starting point for enrichment analysis and experimental follow-up.

\subsection*{Findings from the Extra Trees Classifier (ETC)}

\subsubsection*{A. What was measured and how}

We trained an Extra Trees Classifier (ETC) to distinguish healthy versus T2D single cells using gene-expression features (each gene is a predictor; each cell is an observation). Model evaluation followed a stratified k-fold cross-validation procedure: in each fold the model was fitted on a training split and then tested on held-out cells; out-of-fold predictions were concatenated to obtain unbiased test metrics. Discrimination was summarized with the ROC curve and its area (AUC), while the error profile was summarized with the confusion matrix. Feature importance was computed via mean decrease in impurity (MDI), which ranks genes by how much they reduce class-mixing when used in tree splits.

\subsubsection*{B. Discrimination}

The cross-validated ROC in Fig.~\ref{fig:roc_etc} shows excellent separation between diabetic and healthy cells (mean AUC $\approx 0.999$). This indicates that, across thresholds, the model almost always assigns higher ``diabetic'' scores to T2D cells than to healthy cells. Consistent with this result, similarly high AUC values have been reported for Extra Trees–based models in independent diabetes studies. In particular, Matboli et al.\cite{b22} achieved AUC values close to 0.999 using an Extra Trees Classifier for multi-stage diabetes stratification, supporting the robustness of ETC in capturing complex diabetes-associated biological patterns across different data modalities.

\subsubsection*{C. Error profile (Confusion matrix)}

\begin{figure}
    \centering
    \includegraphics[width=1\linewidth]{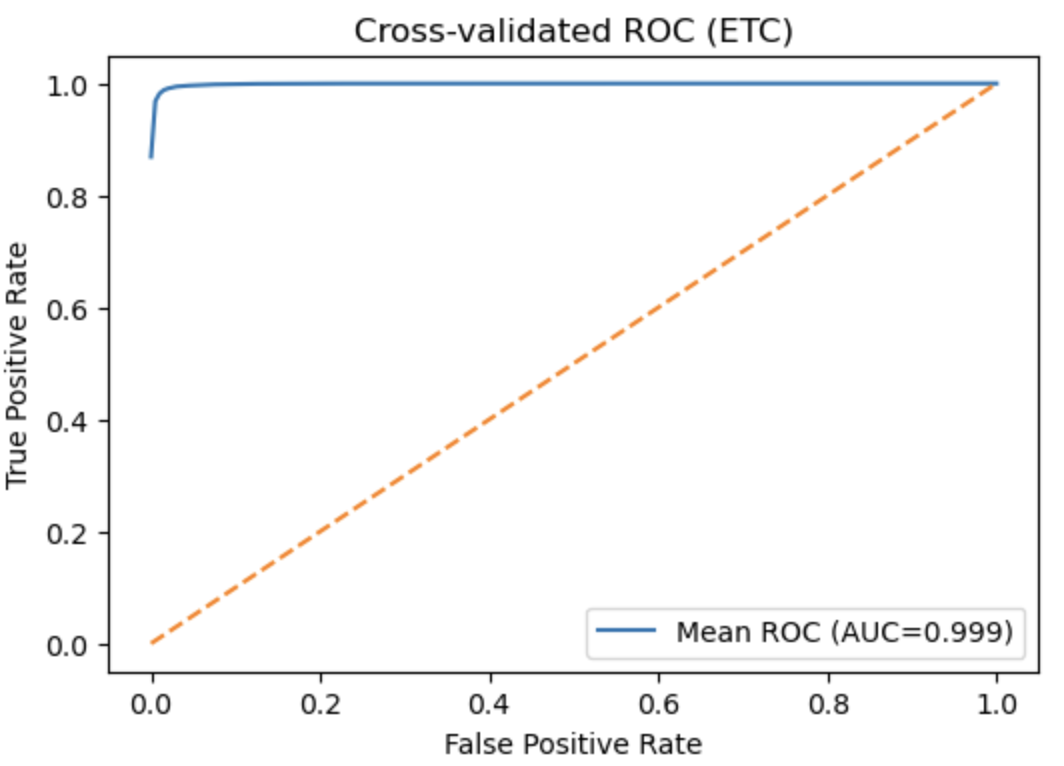}
    \caption{Cross-validated ROC (ETC). The curve lies near the top-left corner; the mean AUC is approximately 0.999.}
    \label{fig:roc_etc}
\end{figure}

From Fig.~\ref{fig:confusion_matrix_etc} the counts are: true negatives = 7{,}876, false positives = 27, false negatives = 204, and true positives = 5{,}416 (total $n = 13{,}523$). These yield overall accuracy $\approx 0.981$, precision (PPV) $\approx 0.995$, recall (sensitivity) $\approx 0.960$, F$_1 \approx 0.978$, and specificity (TNR) $\approx 0.997$. Errors are asymmetric: false negatives (missed T2D cells) are more frequent than false positives (healthy cells flagged as T2D). If recall were the priority (e.g., screening), the decision threshold could be lowered to reduce false negatives at the cost of a modest precision drop, consistent with the ROC trade-off in Fig.~\ref{fig:roc_etc}

\begin{figure}
    \centering
    \includegraphics[width=1\linewidth]{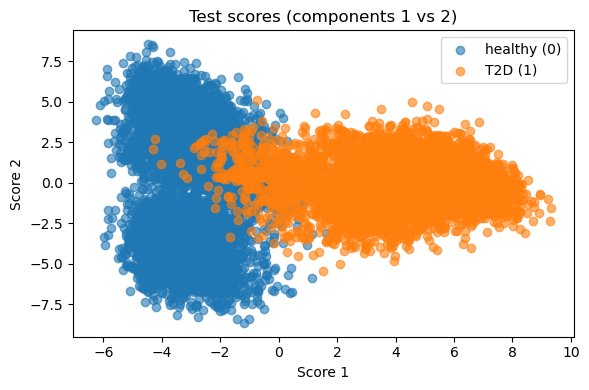}
    \caption{Test scores in a two-component space. Each point is a held-out cell. The healthy (blue) and T2D (orange) clouds are well separated with limited overlap.}
    \label{fig:etc_scores}
\end{figure}

\subsubsection*{D. Score-space visualization}

The 2D score plot in Fig.~\ref{fig:etc_scores} shows two partially separated manifolds: the T2D cluster occupies higher values on ``Score 1'' whereas healthy cells occupy lower values. The limited overlap region corresponds to the confusion-matrix errors and visually explains why sensitivity is slightly lower than precision.

\subsubsection*{E. Gene-level signals (MDI) and biological meaning}

Appendix~II shows the top genes identified by the Extra Trees Classifier (ETC) as most important for distinguishing diabetic from healthy $\beta$-cells. In simple terms, the model highlights which genes contribute most to separating the two conditions. Many of these genes are linked to insulin secretion, stress responses, and cellular organization.

Interestingly, the Vitamin D Binding Protein (Gc) appears again as the top-ranked gene. This does not necessarily imply that it is the single most critical driver of diabetes, but rather that its expression profile shows a strong and consistent difference between healthy and diabetic cells, making it highly informative for the model. Vitamin D signaling has previously been associated with metabolic regulation and immune modulation~\cite{b20}, so its prominence here may reflect broader systemic pathways that intersect with $\beta$-cell health. The prominence of Gc reinforces the possibility that systemic regulators such as vitamin D pathways may also shape $\beta$-cell transcriptional states in diabetes. This supports the view that type\textasciitilde 2 diabetes involves both intrinsic $\beta$-cell dysfunction and broader molecular adaptations to chronic stress.

\begin{figure}
    \centering
    \includegraphics[width=1\linewidth]{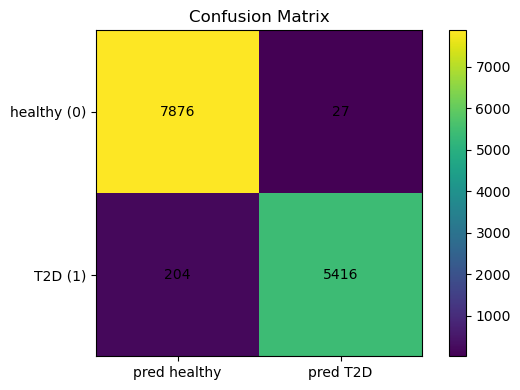}
    \caption{Confusion matrix for the concatenated out-of-fold predictions. Rows are true labels; columns are predicted labels.}
    \label{fig:confusion_matrix_etc}
\end{figure}

\subsubsection*{F. Comparative interpretation (healthy vs.\ T2D)}

These results indicate that T2D cells can be identified with very high discriminative accuracy based on their transcriptomic profiles. The model achieves near-perfect AUC (Fig.~\ref{fig:roc_etc}) and a favourable error profile (Fig.~\ref{fig:confusion_matrix_etc}), with most healthy cells correctly rejected and most T2D cells correctly identified. The remaining errors concentrate in a biologically plausible boundary region (Fig.~\ref{fig:etc_scores}). The gene-importance pattern is consistent with altered insulin processing/secretion and stress adaptation in diabetes, aligning with known beta-cell biology.

\section{Conclusion}

This study set out to explore how single-cell transcriptomic data, when combined with machine learning, can deepen our understanding of the molecular mechanisms underlying type\textasciitilde 2 diabetes (T2D). The motivation was rooted in the recognition that traditional genetic studies, although powerful, often struggle to capture the heterogeneity of cellular responses and the non-linear interactions among thousands of genes. By focusing on the Mouse Islet Atlas, a comprehensive dataset integrating over 300,000 islet cells across developmental stages, stress conditions, and disease states, this work sought to identify transcriptional differences between healthy and diabetic $\beta$-cells and to test whether computational models could reliably classify them.

The application of these machine learning models demonstrated that disease states can be predicted with near-perfect accuracy based solely on transcriptomic profiles. Both models achieved area under the curve (AUC) values approaching 0.999, confirming that the molecular signature of T2D is both strong and consistent. Yet, the models differed in their interpretability and focus. While PLS-DA provided clarity and interpretability, ETC offered robustness, scalability, and slightly higher recall, particularly in identifying diabetic cells.

From a broader perspective, the dual use of PLS-DA and ETC strengthens the validity of the conclusions. The overlap between the two models’ findings, in particular the emphasis on stress-response pathways, insulin secretion machinery, and endocrine signalling, confirms that these molecular programs are robust indicators of diabetic dysfunction. Their convergence also demonstrates how different computational approaches, though methodologically distinct, can complement one another to provide both predictive performance and mechanistic insight.

Beyond model evaluation, this study contributes to the growing evidence that machine learning can serve as a bridge between raw genomic data and biological understanding. By ranking and prioritizing candidate genes, these approaches do not simply produce accurate classifiers but also suggest new leads for experimental validation and potential therapeutic targeting.

At the methodological level, this work illustrates the feasibility of conducting large-scale single-cell transcriptomic analysis in a household laptop, provided that appropriate preprocessing, quality control, and dimensionality reduction are applied. The integration of machine learning further enables the handling of complex, high-dimensional datasets without oversimplifying the underlying biology.

In conclusion, this study shows that combining single-cell transcriptomics with machine learning provides powerful insights into the biology of type\textasciitilde 2 diabetes. It establishes that diabetic $\beta$-cells carry distinct transcriptional fingerprints that can be captured and classified with exceptional accuracy. . It also highlights the complementary strengths of different machine learning methods: PLS-DA offering interpretability and clarity, and ETC delivering robustness and scalability. Together, these approaches underscore the potential of computational biology to accelerate biomarker discovery, refine disease classification, and inform future translational research. Ultimately, this work provides a proof-of-concept for how integrative, data-driven approaches can transform our understanding of complex diseases such as type\textasciitilde 2 diabetes, bridging the gap between genomic information and clinical application.

\appendices
\onecolumn
\section{Top genes ranked by VIP scores}

Appendix~A reports the top-ranked genes identified by the Partial Least Squares Discriminant Analysis (PLS-DA) model using Variable Importance in Projection (VIP) scores. VIP scores indicate how strongly each gene contributes to the components that separate diabetic from healthy $\beta$-cells. Genes with higher VIP values have a greater impact on class discrimination and are therefore considered highly informative for distinguishing type~2 diabetic from non-diabetic gene expression profiles.

\begin{table}[htbp]
\centering
\caption{Top genes ranked by VIP scores from the PLS-DA model.}
\label{tab:vip_genes_plsda}
\resizebox{\columnwidth}{!}{%
\begin{tabular}{l c l c c c c}
\hline
\textbf{Gene ID} & \textbf{VIP} & \textbf{Gene name} & \textbf{Entrez} & \textbf{GO BP} & \textbf{GO MF} & \textbf{KEGG} \\
\hline
ENSMUSG00000035540 & 5.1438 & Vitamin D binding protein & 14473 & -- & -- & -- \\
ENSMUSG00000026335 & 4.8293 & Peptidylglycine alpha-amidating monooxygenase & 18484 & -- & -- & -- \\
ENSMUSG00000032181 & 4.7204 & Secretogranin III & 20255 & -- & protein binding & -- \\
ENSMUSG00000022490 & 4.6582 & Protein phosphatase 1 regulatory inhibitor & 58200 & -- & -- & -- \\
ENSMUSG00000003355 & 4.6332 & FK506 binding protein 11 & 66120 & -- & -- & -- \\
ENSMUSG00000050711 & 4.6181 & Secretogranin II & 20254 & -- & -- & -- \\
ENSMUSG00000032532 & 4.5251 & Cholecystokinin & 12424 & -- & -- & -- \\
ENSMUSG00000009246 & 4.5148 & TRP cation channel, subfamily M & 56843 & -- & -- & -- \\
ENSMUSG00000015401 & 4.5067 & Collectrin & 57394 & -- & -- & -- \\
ENSMUSG00000018451 & 4.4713 & RIKEN cDNA 6330403K07 & 103712 & -- & -- & -- \\
ENSMUSG00000031762 & 4.4469 & Metallothionein 2 & 17750 & -- & -- & -- \\
ENSMUSG00000015134 & 4.3395 & Aldehyde dehydrogenase 1A3 & 56847 & -- & -- & -- \\
ENSMUSG00000026989 & 4.3109 & Death associated protein-like 1 & 76747 & -- & -- & -- \\
ENSMUSG00000044139 & 4.2445 & Serine protease 53 & 330657 & proteolysis & -- & -- \\
ENSMUSG00000031271 & 4.2122 & Serine peptidase inhibitor (clade) & 331535 & -- & -- & -- \\
ENSMUSG00000023236 & 4.1112 & Secretogranin V & 20394 & -- & -- & -- \\
ENSMUSG00000022315 & 4.1024 & Solute carrier family 30 & 239436 & -- & -- & -- \\
ENSMUSG00000034871 & 4.0958 & FAM151A & 230579 & biological\_process & molecular\_function & -- \\
ENSMUSG00000037706 & 4.0094 & CD81 antigen & 12520 & -- & -- & -- \\
ENSMUSG00000045763 & 3.9721 & Brain abundant membrane signal protein & 70350 & -- & -- & -- \\
ENSMUSG00000096956 & 3.9347 & Small nucleolar RNA host gene 18 & 100616095 & biological\_process & molecular\_function & -- \\
\hline
\end{tabular}}
\end{table}

\section{Top-ranked genes identified by the Extra Trees Classifier}

Appendix~B presents the top-ranked genes identified by the Extra Trees Classifier (ETC) based on the Mean Decrease in Impurity (MDI) measure. MDI quantifies the average reduction in node impurity achieved when a given gene is used to split the data across all trees in the ensemble, weighted by the proportion of samples reaching each split. Genes with higher MDI values therefore contribute more consistently to improving class separation, indicating that their expression patterns are particularly informative for distinguishing healthy from diabetic $\beta$-cells during model training.

\begin{table}[htbp]
\centering
\caption{Top-ranked genes identified by the Extra Trees Classifier (ETC) using mean decrease in impurity (MDI).}
\label{tab:etc_genes_mdi}
\resizebox{\columnwidth}{!}{%
\begin{tabular}{l c l l c l l}
\hline
\textbf{Gene} & \textbf{Importance} & \textbf{Symbol} & \textbf{Name} & \textbf{Entrez} & \textbf{GO BP} & \textbf{GO MF} \\
\hline
Gc & 0.028543 & Gc & vitamin D binding protein & 14473 & -- & -- \\
Cck & 0.024660 & Cck & cholecystokinin & 12424 & -- & -- \\
6330403K07Rik & 0.020999 & 6330403K07Rik & RIKEN cDNA 6330403K07 gene & 103712 & -- & -- \\
Ppp1r1a & 0.020508 & Ppp1r1a & protein phosphatase 1, regulatory inhibitor subunit 1A & 58200 & -- & -- \\
Fkbp11 & 0.019680 & Fkbp11 & FK506 binding protein 11 & 66120 & -- & -- \\
Trpm5 & 0.019238 & Trpm5 & transient receptor potential cation channel, subfamily M, member 5 & 56843 & -- & -- \\
Pam & 0.018175 & Pam & peptidylglycine alpha-amidating monooxygenase & 18484 & -- & -- \\
Aldh1a3 & 0.017711 & Aldh1a3 & aldehyde dehydrogenase family 1, subfamily A3 & 56847 & -- & -- \\
Mt2 & 0.016441 & Mt2 & metallothionein 2 & 17750 & -- & -- \\
Cd81 & 0.016030 & Cd81 & CD81 antigen & 12520 & -- & -- \\
Fam151a & 0.015783 & Fam151a & family with sequence similarity 151, member A & 230579 & biological\_process & molecular\_function \\
Serpina7 & 0.015044 & Serpina7 & serine peptidase inhibitor, clade A, member 7 & 331535 & -- & -- \\
Cltrn & 0.014270 & Cltrn & collectrin, amino acid transport regulator & 57394 & -- & -- \\
Prss53 & 0.014048 & Prss53 & serine protease 53 & 330657 & proteolysis & -- \\
Dapl1 & 0.013870 & Dapl1 & death associated protein-like 1 & 76747 & -- & -- \\
Scg3 & 0.012931 & Scg3 & secretogranin III & 20255 & -- & protein binding \\
Slc30a8 & 0.012821 & Slc30a8 & solute carrier family 30 (zinc transporter), member 8 & 239436 & -- & -- \\
Xist & 0.012600 & Xist & inactive X specific transcripts & 213742 & -- & -- \\
Pappa2 & 0.011884 & Pappa2 & pappalysin 2 & 23850 & -- & -- \\
Scg2 & 0.011875 & Scg2 & secretogranin II & 20254 & -- & -- \\
\hline
\end{tabular}}
\end{table}

\vspace{12pt}

\end{document}